\documentstyle[12pt]{article}

\def\brbsg{B(b\to s\gamma)}
\def\bsg{b\to s\gamma}
\def\Zbb{Z\rightarrow b\ov b}
\def\beq{\begin{equation}}
\def\eeq{\end{equation}}

\catcode`\@=11 % This allows us to modify PLAIN macros.
\def\coeff#1#2{{\textstyle{#1\over #2}}}

\def\lsim{\mathrel{\mathpalette\@versim<}}
\def\gsim{\mathrel{\mathpalette\@versim>}}
\def\@versim#1#2{\vcenter{\offinterlineskip
    \ialign{$\m@th#1\hfil##\hfil$\crcr#2\crcr\sim\crcr } }}
\def\etal{{\em et. al.}}
\def\JL{J. L. Lopez}
\def\DVN{D. V. Nanopoulos}

\def\t1{{\tilde 1}}
\def\ov{\overline}

\def\GeV{\,{\rm GeV}}

\def\to{\rightarrow}

\def\NPB#1#2#3{Nucl. Phys. B {\bf#1} (19#2) #3}
\def\PLB#1#2#3{Phys. Lett. B {\bf#1} (19#2) #3}

\def\PRD#1#2#3{Phys. Rev. D {\bf#1} (19#2) #3}
\def\PRL#1#2#3{Phys. Rev. Lett. {\bf#1} (19#2) #3}

\def\TAMU#1{Texas A \& M University preprint CTP-TAMU-#1}

\begin{document}
% TH format
\begin{flushright}
\baselineskip=12pt
%{CERN-TH.????/93}\\
{CTP-TAMU-54/93}\\
{ACT-21/93}\\
\end{flushright}
% PPE format
%\begin{center}
%{\large EUROPEAN ORGANIZATION FOR NUCLEAR RESEARCH}
%\end{center}
%\begin{flushright}
%{CERN-PPE/93-??}\\
%{?? July, 1993}\\
%{CERN-LAA/93-??}\\
%{CERN-TH.????/93}\\
%{CTP-TAMU-40/93}\\
%{ACT-15/93}\\
%\end{flushright}
%

\begin{center}
\vglue 0.3cm
{\Large\bf $b\rightarrow s\gamma$ and $Z\rightarrow b\ov b$ Constraints on\\}
\vspace{0.2cm}
{\Large\bf Two Higgs Doublet Model\\}
\vglue 1.5cm
{GYE T. PARK\\}
\vglue 0.4cm
{\em Center for Theoretical Physics, Department of Physics, Texas A\&M
University\\}
{\em College Station, TX 77843--4242, USA\\}
{\em and\\}
{\em Astroparticle Physics Group, Houston Advanced Research Center
(HARC)\\}
{\em The Woodlands, TX 77381, USA\\}
\baselineskip=12pt

\vglue 2.0cm
{\tenrm ABSTRACT}
\end{center}
\vglue -0.2cm
{\rightskip=3pc
 \leftskip=3pc
%\xpt\baselineskip=12pt
\noindent
We perform a combined analysis of two stringent constraints on two Higgs
doublet model,
coming from the recently announced CLEO II bound on $\brbsg$ and from the
recent LEP data on the ratio $\Gamma(Z\rightarrow b\ov
b)\over{\Gamma(Z\rightarrow hadrons)}$.
We include one-loop vertex corrections to $\Zbb$ in the model.
We find that although the CLEO II
bound serves as the strongest constraint present in the charged Higgs sector of
the model, the current
LEP value for $R_b$ may also provide a further constraint for $\tan\beta<1$.
}
% TH format
\vspace{6cm}
\begin{flushleft}
\baselineskip=12pt
%{CERN-TH.????/93}\\
{CTP-TAMU-54/93}\\
{ACT-21/93}\\
August 1993
\end{flushleft}
\vfill\eject
\setcounter{page}{1}
\pagestyle{plain}

\baselineskip=14pt

%\section{Introduction}
Despite the remarkable successes of the Standard Model(SM) in its complete
agreement with current all experimental data, there is still no
experimental information on the nature of its Higgs sector.
The 2 Higgs doublet model(2HDM) is one of the mildest extensions of the SM,
which has been consistent with experimental data.
In the 2HDM to be considered here, the Higgs sector consists of 2 doublets,
$\phi_1$ and $\phi_2$,
 coupled to the charge -1/3 and +2/3 quarks, respectively, which will ensure
the absence of Flavor-Changing Yukawa couplings at the tree level
 \cite{NOFC}. The physical Higgs spectrum of the model includes two CP-even
 neutral Higgs($H^0$, $h^0$), one CP-odd neutral Higgs($A^0$)
, and a pair of charged Higgs($H^\pm$). In addition to the masses of these
Higgs, there is another free parameter in the model, which is $\tan\beta\equiv
v_2/v_1$, the ratio of the vacuum expectation values of both doublets.

With a revived interest on the
flavor-changing-neutral-current (FCNC) $\bsg$ decay, spurred by the
CLEO bound $\brbsg<8.4\times10^{-4}$ at $90\%$ C.L. \cite{CLEO}, it was pointed
out recently that the CLEO bound can be violated due to the charged Higgs
contribution in the 2HDM and the Minimal Supersymmetric Standard Model(MSSM)
basically if $m_{H^\pm}$ is too light, excluding large portion of the charged
Higgs parameter space \cite{BargerH}.
The recently announced CLEO II bound $\brbsg<5.4\times10^{-4}$ at
$95\%$\cite{Thorndike} excludes even larger portion of the parameter space
\cite{VernonHARC}. It has certainly proven that this particular decay mode can
provide more stringent constraint on new physics beyond SM than any other
experiments\cite{bsgamma}.
In this report, we will show that in addition to the CLEO II bound on $\bsg$,
 one may be able to constrain the 2HDM further, by incorporating one-loop
vertex corrections to $Z\rightarrow b\ov b$,
 with the recent LEP data on $R_b$\cite{LP93}, which is defined to be the ratio
 $\Gamma(Z\rightarrow b\ov b)\over{\Gamma(Z\rightarrow hadrons)}$.
A number of authors in fact have studied the one-loop corrections to
$\Gamma(Z\rightarrow b\ov b)$
in the context of the SM\cite{RbSM} and a few extensions of it
\cite{Rb2HD,BF,Denner} although the emphases were different.

In the 2HDM, $\bsg$ decay receives contributions from penguin diagrams with
$W^\pm-t$ loop and $H^\pm-t$ loop.
The expression used for $\brbsg$ is given by \cite{BG}
\beq
{B(b\to s\gamma)\over B(b\to ce\bar\nu)}={6\alpha\over\pi}
{\left[\eta^{16/23}A_\gamma
+\coeff{8}{3}(\eta^{14/23}-\eta^{16/23})A_g+C\right]^2\over
I(m_c/m_b)\left[1-\coeff{2}{3\pi}\alpha_s(m_b)f(m_c/m_b)\right]},
\eeq
where $\eta=\alpha_s(M_Z)/\alpha_s(m_b)$, $I$ is the phase-space factor
$I(x)=1-8x^2+8x^6-x^8-24x^4\ln x$, and $f(m_c/m_b)=2.41$ the QCD
correction factor for the semileptonic decay.
We use the 3-loop expressions for $\alpha_s$ and choose $\Lambda_{QCD}$ to
obtain $\alpha_s(M_Z)$ consistent with the recent measurements at LEP.
In our computations we have used: $\alpha_s(M_Z)=0.118$, $ B(b\to
ce\bar\nu)=10.7\%$, $m_b=4.8\GeV$, and
$m_c/m_b=0.3$. The $A_\gamma,A_g$ are the
coefficients of the effective $bs\gamma$ and $bsg$ penguin operators
evaluated at the scale $M_Z$. The contributions to $A_{\gamma ,g}$ from the
$W^\pm-t$ loop, the $H^\pm-t$ loop are given in Ref\cite{BG}.
As mentioned above, the CLEO II bound excludes a large portion of the parameter
space. In Fig. 1 we present the excluded regions in ($m_{H^\pm}$,
$\tan\beta$)-plane for $m_t=120, 130$, and $150\GeV$, which lie to the left of
each curve. We have imposed in the figure also the lower bound on $\tan\beta$
from  ${m_t\over{600}}\lsim\tan\beta\lsim{600\over{m_b}}$ obtained by demanding
that the theory remain perturbative\cite{BargerLE}.
We see from the figure that at large $\tan\beta$ one can obtain a lower bound
on $m_{H^\pm}$ for each value of $m_t$. And we obtain the bounds
, $m_{H^\pm}\gsim 160, 186, 244\GeV$ for $m_t=120, 130, 150\GeV$, respectively.
Our strategy is now to sample $\tan\beta$ and $m_{H^\pm}$ from the allowed
regions in Fig. 1,  which will be in turn
used below to calculate $R_b$ at one-loop level in this model.

In the SM, the diagrams for the vertex corrections to $\Zbb$ involve
top quarks and $W^\pm$ bosons. However, in the 2HDM there are additional
diagrams involving $H^\pm$ bosons instead of $W^\pm$ bosons. These additional
diagrams
have been calculated in Ref\cite{Rb2HD,BF,Denner}. Here we use the formulas
given in Ref\cite{BF}. The calculation involves numerical evaluation of the
reduced Passarino-Veltman functions\cite{Ahn}.
Our numerical results agree with those in Fig. 4-5 in Ref\cite{BF}.
In our calculation, we neglect  the neutral Higgs contributions which are all
proportional to $m_b^2\tan\beta^2$ and
become sizable only for
$\tan\beta>{m_t\over{m_b}}$ and very light neutral Higgs $\lsim50\GeV$, but
decreases rapidly to get negligibly small as the Higgs masses become
$\gsim100\GeV$\cite{Denner}.
We also neglect oblique corrections from the Higgs bosons just to avoid
introducing more paramters. This correction grows as $m^2_{H^\pm}$ if
$m_{H^\pm}\gg m_{H^0,h^0,A^0}$, and gives only a $-0.1\%$ correction for
$m_{H^\pm}=500\GeV$.
Although $\tan\beta\gg1$ seems more appealing because of apparent hierarchy
 $m_t\gg m_b$, there are still no convincing arguments against $\tan\beta<1$.
 Here we choose to explore the region of $\tan\beta\lsim1$.
In Fig. 2 we show the model predictions for $R_b$ as a function of $m_t$ in
comparison with the SM prediction. The parameters were sampled  from the
allowed regions in the Fig. 1. Two horizontal solid lines in the figure
represent the lower limits from the recent LEP data $R_b=0.2203\pm
0.0027$\cite{LP93}. The upper one corresponds to the 1-$\sigma$ value, and the
lower one to the 1.64-$\sigma$ value. We note that the deviations from the SM
can be quite large for $\tan\beta<1$ because the charged Higgs contribution
grows as
$m^2_t/\tan^{2}\beta$ for $\tan\beta\ll{m_t\over{m_b}}$. The deviations for
$\tan\beta<1$ can be as large as $-2.2\%$ for $m_t=150\GeV$ while they become
much smaller for $\tan\beta>1$\cite{Bigtanb}.
We have also considered other constraints from low-energy data primarily in
$B-\ov{B}, D-\ov{D}, K-\ov{K}$ mixing that exclude low values of
$\tan\beta$\cite{BargerLE,LowEdata}. But $\tan\beta\lsim0.5$ seems to be still
allowed by these constraints for $m_t\lsim150\GeV$ and $m_{H^\pm}\gsim250\GeV$.
However, for small values
 of $\tan\beta$, as is seen from Fig. 2, the model predicts $m_t\lsim110\GeV$
at 1-$\sigma$ level, which is in conflict with the recent bound from
CDF\cite{CDFmt}, $m_t>113\GeV$ although the corresponding prediction at
1.64-$\sigma$ level is still allowed .
These low values of $\tan\beta\lsim 0.5$ might be at the verge of being
disfavored
even at $90\%$C.L. as the experimental lower bound on $m_t$ tends to grow.
Nevertheless, the CLEO II
 bound is still by far the strongest constraint present in the charged Higgs
sector of the model especially for $\tan\beta\gsim 1$.

In the MSSM, the situation becomes much more complicated.
In the calculation of $\brbsg$ in Ref\cite{BargerH}, other important
contributions such as chargino-squark contribution were not included.
However, it was shown very recently\cite{BG} that the bound on $m_{H^\pm}$ in
Ref
\cite{BargerH} could well be evaded in the full supersymmetric
calculation since the $\bsg$ amplitude
vanishes in the exact supersymmetric limit. Therefore, with full calculations,
one may get significantly smaller
bounds on $m_{H^\pm}$ than the ones given above\cite{bsgamma}. However, adding
supersymmetric
particles raises $R_b$ significantly above the SM value due to chargino-squark
loops and neutralino-squark loops cancelling out to a great extent the charged
Higgs
constribution and also the standard contribution\cite{BF}. Thus, this makes
it extremely difficult for one to be able to constrain the MSSM with the
current LEP data.

In conclusion, we have performed a combined analysis of two stringent
constraints
on the 2 Higgs doublet model, coming from the recently announced  CLEO II bound
on $\brbsg$ and from the recent LEP data on $R_b$.
We have included one-loop vertex corrections to $\Zbb$ in the model.
We find that although the CLEO II
bound serves as the strongest constraint present in the charged Higgs sector of
the model, the current
LEP value for $R_b$ may also provide a further constraint for $\tan\beta<1$.

\section*{Acknowledgements}
The author thanks the Center for Theoretical Physics at Seoul National
University for their hospitality during early stage of this work.
This work has been supported by the World Laboratory.

\newpage
\noindent{\large\bf Figure Captions}
\begin{description}
\item Figure 1: The regions in $(m_{H^\pm},\tan\beta)$ plane excluded by
the CLEO II bound $\brbsg<5.4\times10^{-4}$ at $95\%$, for $m_t=120, 130, 150
\GeV$ in 2HDM. The excluded regions
lie to the left of each curve. The values of $m_t$ used are as indicated.
\item Figure 2:  $R_b$ as a function of top mass in the SM(Solid), and the
2HDM for five different sets of $\tan\beta$ and $m_{H^\pm}$ sampled from the
allowed
regions in the Fig. 1. The values of ($\tan\beta$, $m_{H^\pm}$) used  are as
indicated under each curve. Two horizontal solid lines represent the lower
limits from the recent LEP data $R_b=0.2203\pm 0.0027$. The upper one
corresponds to the 1-$\sigma$ value, and the lower one to the 1.64-$\sigma$
value. Anything above these lines is allowed.
\end{description}


\begin{thebibliography}{99}
\bibitem{Bigtanb} For $\tan\beta>{m_t\over{m_b}}$, the charged Higgs
contribution can also become sizable due to its large $\tan\beta$ dependence of
$m^2_b\tan^2\beta$. For example, it yields a $-0.25\%$ deviation for
$\tan\beta=70$, $m_{H^\pm}=250\GeV$, and $m_t=150\GeV$. For large $\tan\beta$,
see Ref\cite{Denner}.
\end{thebibliography}
\end{document}